\documentclass[a4paper]{PoS}
\usepackage{epsfig}
\usepackage{bm}
\usepackage{amssymb}
\usepackage{fontenc}
\usepackage{times}
\usepackage{mathptmx}
\usepackage{graphicx}

\title{Phase Quenched Lattice QCD at Finite Density and Temperature}

\ShortTitle{Phase Quenched Lattice QCD at Finite Density and Temperature}

\author{\speaker{D.~K.~Sinclair} \\
        HEP Division, Argonne National Laboratory, 9700 South Cass Ave.,
        Argonne, IL, 60439, USA 
        \\
        E-mail: \email{dks@hep.anl.gov}}

\author{J.~B.~Kogut\\
Department of Energy, Division of High Energy Physics, Washington, DC 20585,
USA\\
and\\
Dept. of Physics -- TQHN, Univ. of Maryland, 82 Regents Dr., College
Park, MD 20742, USA\\
        E-mail: \email{jbkogut@umd.edu}}

\abstract{We simulate 3-flavour lattice QCD at finite quark-number chemical
potential $\mu$ in the phase-quenched approximation, close to the finite
temperature transition. Working close to the critical quark mass, we find no
evidence for the expected critical endpoint at small $\mu$. We are performing 
further simulations aimed at calculating the equation-of-state of this theory 
outside of the superfluid domain, where its phase structure is expected to
mimic the full theory.}

\FullConference{The XXV International Symposium on Lattice Field Theory\\
		 July 30 - August 4 2007\\
		 Regensburg, Germany}

\begin{document}

\section{Introduction}

QCD at finite chemical potential $\mu$ has a complex fermion determinant, which
prevents direct application of the standard simulation methods of lattice QCD.
We adopt the phase-quenched approximation, where we replace the fermion
determinant by its magnitude, which allows the use of standard simulation
methods. Outside of the superfluid phase in the $(T,\mu)$ plane, it is likely
that the phase structures of full and phase-quenched QCD are the same.
We use exact RHMC simulations \cite{Clark:2006wp}, employing a speculative 
lower bound to the spectrum of the Dirac operator to enable the required use of
rational approximations \cite{Kogut:2006jg}.

We simulate 3-flavour lattice QCD on $8^3 \times 4$, $12^3 \times 4$ and
$16^3 \times 4$ lattices for masses near to the critical mass $m_c$ at $\mu=0$, 
for temperatures close to the transition temperature from hadronic matter to a 
quark-gluon plasma. If the critical mass increases with increasing $\mu$, 
then for masses just above $m_c(0)$ we would find a critical endpoint at
$m=m_c(\mu)$ for small $\mu$. However, as our simulations indicate, $m_c$
decreases with increasing $\mu$ and no critical endpoint is found for small
$\mu$ \cite{Sinclair:2006zm}. 
We introduce $\mu_I=2\mu$, which has the interpretation of an isospin
chemical potential (at least for even numbers of flavours).

We are now simulating this same 3-flavour lattice QCD for a range of $\mu$
and $\beta$ values outside the superfluid phase, in order to calculate the
equation-of-state for phase-quenched QCD.

\section{Simulations and results}

The nature of the transition is best determined on finite lattices using the
4th-order Binder cumulant for the magnetic order parameter. Since, at finite
quark mass, we do not know this order parameter, we use the chiral condensate,
introducing (hopefully small) finite size effects. The Binder cumulant for any
observable $X$ is defined by \cite{Binder:1981sa}
\begin{equation}
B_4(X) = {\langle(X-\langle X \rangle)^4\rangle \over
          \langle(X-\langle X \rangle)^2\rangle^2} .
\end{equation}
If there is a critical endpoint at small $\mu_I$ for $m > m_c(0)$, then
the Binder cumulant should decrease from its crossover value $B_4=3$, passing
through the Ising value $B_4=1.604(1)$ at the endpoint, and falling towards
its first-order value $B_4=1$ as $\mu_I$ is increased. Small $\mu_I$ means
$\mu_I$ small enough to lie outside of the superfluid phase. For $T=0$, this
means $\mu_I < m_\pi$.

Since, as determined in these simulations, $m_c(\mu_I=0)=0.0265(3)$, we 
perform simulations at quark mass $m=0.02$, $m=0.025$, $m=0.03$ and $m=0.035$.
At the 3 larger masses we simulate at $\mu_I=0$, $\mu_I=0.2$, $\mu_I=0.3$,
while for $m=0.02$ we only simulate at $\mu_I=0$. On the $12^3 \times 4$
lattice where we have the highest statistics, we generate 300,000 trajectories
at each of $4$ $\beta$ values close to $\beta_c$ (the transition value), and
use Ferrenberg-Swendsen reweighting \cite{Ferrenberg:1988yz} to continue to 
$\beta_c$, taken as the $\beta$ which minimizes $B_4(\bar{\psi}\psi)$. Five
noisy estimators of $\bar{\psi}\psi$ per trajectory are used to obtain an
unbiased estimator for $B_4$.

\begin{figure}[htb]
\epsfxsize=4in
\centerline{\epsffile{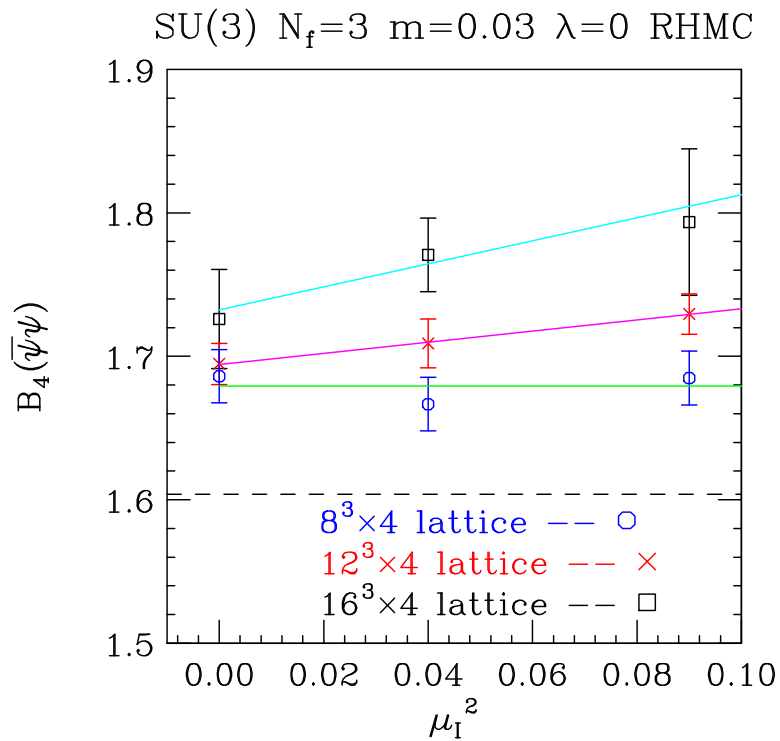}}
\caption{Binder cumulants at $T=T_c$ as a functions of $\mu_I^2$ at $m=0.030$.}
\label{fig:B4(m0.03)}
\end{figure}

Figure~\ref{fig:B4(m0.03)} shows these Binder cumulants as functions of $\mu_I$
for $m=0.03$ on $8^3 \times 4$, $12^3 \times 4$ and $16^3 \times 4$ lattices.
Rather than decrease with increasing $\mu_I$, the curves for the larger lattices
show a modest increase with increasing $\mu_I$, and hence no sign of a critical
endpoint. The graphs for $m=0.035$ are similar, except that we do not have 
`data' on $16^3 \times 4$ lattices. At $m=0.025$, although there is a 
suggestion of such an increase, but the `data' is consistent with no $\mu_I$
dependence.

\begin{figure}[htb] 
\epsfxsize=4in      
\centerline{\epsffile{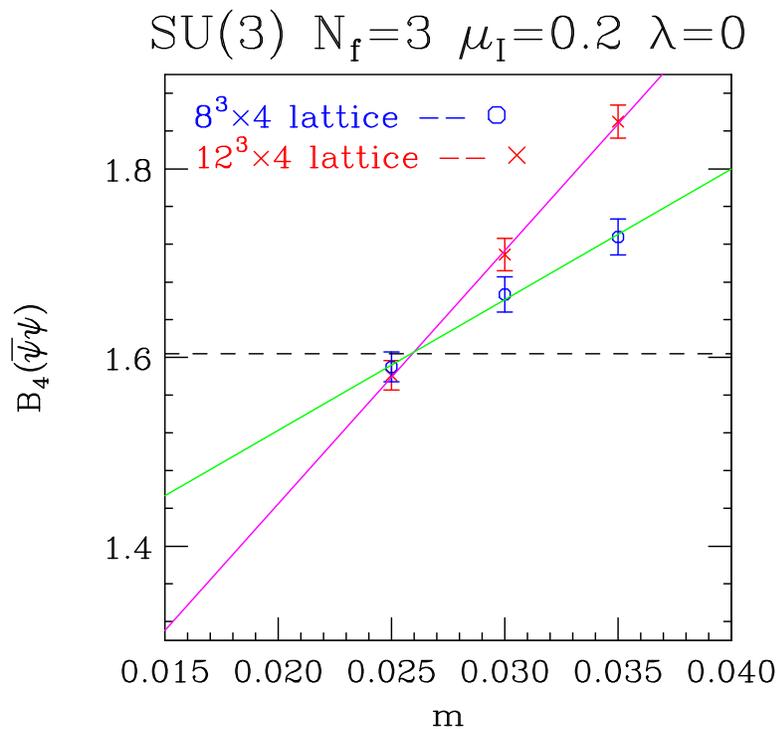}}
\caption{Binder cumulants at $T=T_c$ as a functions of $m$ at $\mu_I=0.2$.}
\label{fig:B4(mu0.2)}                                                         
\end{figure} 

It is instructive to examine the behaviour of $B_4$ as functions of $m$ at
fixed $\mu_I$ for evidence of a critical point in the universality class of
the 3-dimensional Ising model. We find that for each of our 3 $\mu_I$ values
the curves for the $8^3 \times 4$ and $12^3 \times 4$ lattices cross very
close to the Ising value. Not only does this indicate that the critical point
belongs to the Ising universality class, but also suggests that using 
$\bar{\psi}\psi$ as the magnetic order parameter is a reasonable choice.
Figure~\ref{fig:B4(mu0.2)} shows $B_4$ as functions of $m$ for $\mu_I=0.2$.
Estimating the position of the critical points as the values of $m$ where the
$12^3 \times 4$ curves achieve their Ising value gives: $m_c(0)=0.0265(3)$,
$m_c(0.2)=0.0259(5)$ and $m_c(0.3)=0.0256(4)$, i.e. $m_c(\mu_I)$ is a slowly
decreasing function of $\mu_I$.

\begin{figure}[htb]
\epsfxsize=4in
\centerline{\epsffile{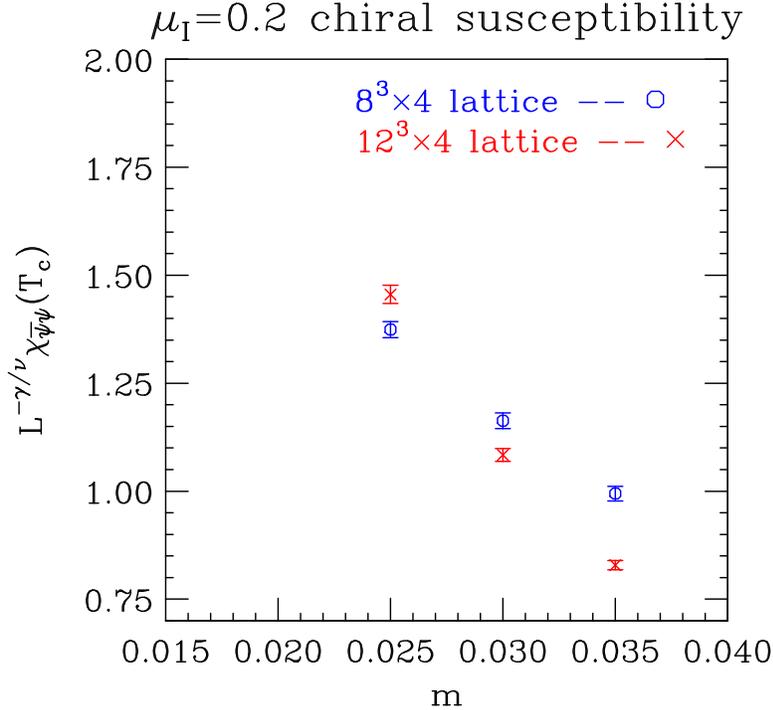}}
\caption{Finite size scaling for the peak of the chiral susceptibilities at
$\mu_I=0.2$}
\label{fig:chi(mu0.2)}
\end{figure}

From the same `data' we have also calculated the chiral susceptibility
\begin{equation}                   
\chi_{\bar{\psi}\psi} = {V \over T} \langle\langle\bar{\psi}\psi^2\rangle
                                   -\langle\bar{\psi}\psi\rangle^2\rangle
\end{equation}
where the $\bar{\psi}\psi$s on the right-hand side are lattice averaged
quantities. Finite size scaling at the critical point predicts that
\begin{equation}
\chi_{\bar{\psi}\psi}(L,T_c) = L^\frac{\gamma}{\nu} \tilde{\chi}.
\end{equation}
Hence if we plot $L^{-\frac{\gamma}{\nu}}\chi_{\bar{\psi}\psi}(L,T_c)$ 
as functions of $m$ for different $L$s, the curves should intersect at the
critical point. In figure~\ref{fig:chi(mu0.2)} we plot this quantity for 
$\mu_I=0.2$, and note that curves for $8^3 \times 4$ and $12^3 \times 4$
lattices would intersect between $m=0.025$ and $m=0.03$ which is where we
found the critical point from the Binder cumulants. Similar results obtain
for $\mu_I=0$ and $\mu_I=0.3$.

\section{Equation of state}

\begin{figure}[htb]
\epsfxsize=4in
\centerline{\epsffile{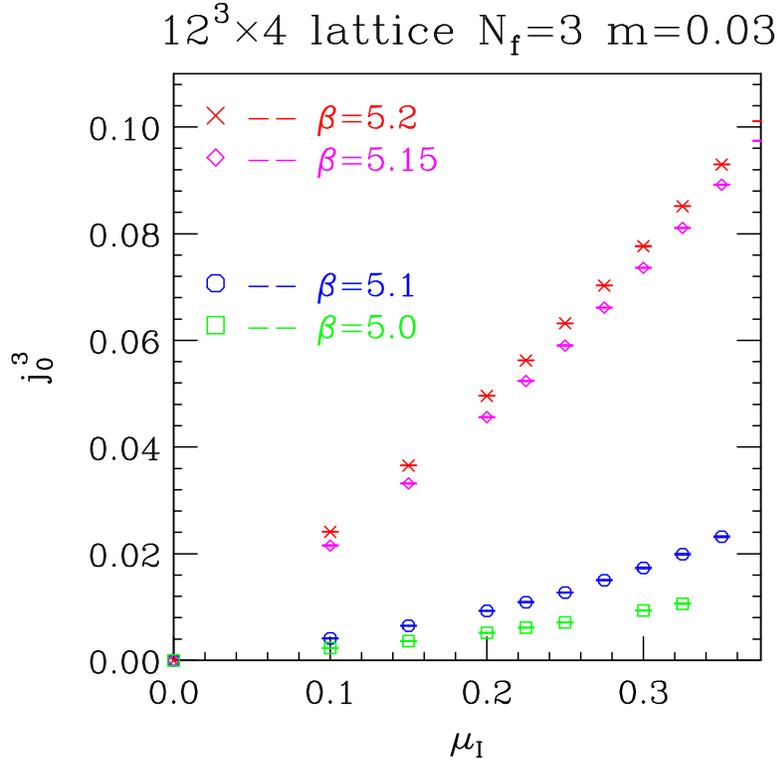}}
\caption{Isospin density as functions of $\mu_I$ at fixed $\beta$ values}
\label{fig:j03}
\end{figure}

The equation-of-state (EOS) expresses the pressure $p$, the entropy density $s$
and the energy density $\epsilon$ as functions of temperature $T$ and $\mu_I$.
(Calculations of the equation of state for QCD at finite $T$ and $\mu$ have
been performed for example in \cite{Ejiri:2005uv} and \cite{Csikor:2004ik}.)
The pressure $p$ is simply related to the partition function through:
\begin{equation}
p = {T \over V}\ln Z(T,\mu_I).
\end{equation}
However, we do not actually measure the partition function $Z$ in our
simulations, only observables. $Z(T,0)$ can be calculated by numerically
integrating
\begin{equation}
{d \ln Z \over d \beta} = \langle 6 {V \over T} S_g \rangle
\end{equation}
where $S_g$ is the plaquette action. We can then numerically integrate
\begin{equation}
{d \ln Z \over d \mu_I} = \langle {N_f \over 8}{V \over T} j_0^3 \rangle
\end{equation}
at constant $\beta$, where $\frac{N_f}{8}j_0^3$ is the the isospin density, to
obtain $Z(T,\mu_I)$. Figure~\ref{fig:j03} shows the $\mu_I$ dependence of
$j_0^3$ at fixed $\beta$ values. For the upper 2 $\beta$s we also have `data'
for larger $\mu_I$s, up to saturation which occurs for $\mu_I \approx 2$.

To obtain $T$ in physical units requires knowledge of the running of the
coupling constant, $\beta=\beta(a)$. This is determined at $\mu_I=0$.
Once this running of the coupling constant is known, this can be used to
determine $\epsilon$, since
\begin{equation}
\epsilon = {T^2 \over V} {\partial \over \partial T} \ln Z .
\end{equation}

\section{Discussion and conclusions}

For $\mu$ small enough or $T$ large enough that phase-quenched QCD is in its
normal rather than its superfluid phase, full QCD and phase-quenched QCD are
expected to have the same phase structure. Lattice simulations indicate that
the fluctuations of the phase of the fermion determinant, on lattices large
enough to observe this phase structure, are sufficiently small for this to be
so \cite{Allton:2002zi}. 
Random matrix calculations in the epsilon regime agree with this conclusion
\cite{Splittorff:2007ck}.

For 3-flavour lattice QCD our phase-quenched simulations show no evidence for
a critical endpoint in the range of $m$ and $\mu$ values where it would have
been expected if it were associated with the critical point observed when the
quark mass is varied. This is in agreement with the results of de Forcrand and
Philipsen obtained using analytic continuation methods from simulations at
imaginary $\mu$, for full QCD \cite{de Forcrand:2006pv}.

We suggest that the softening of the transition as $\mu_I$ is increased is
because the introduction of an isospin chemical potential reduces the symmetry
(at least for even numbers of flavours). Reducing symmetry tends to soften
transitions. For example, reducing the number of flavours from 3 to 2, reduces
the symmetry at $m=0$, $\mu=0$ from $SU(3) \times SU(3)$ to $SU(2) \times
SU(2)$ and the transition softens from first order to second. The addition of
a mass breaks chiral symmetry to a pure vector symmetry, and softens phase
transitions to crossovers.

Determination of the true scaling field for the magnetic order parameter --
a linear combination of the chiral condensate, the plaquette action and the
isospin density -- is needed to remove the finite size effects. Use of the
methods of de Forcrand and Philipsen to improve the signal/noise in calculating
the $\mu^2$ dependence of the Binder cumulants, would also help 
\cite{deforcrand,philipsen}.

Fodor and Katz have predicted a critical endpoint at $T_E = 162(2)$~MeV and 
$\mu =120(13)$~MeV, using their reweighting methods \cite{Fodor:2004nz}. 
However, this $\mu$ is
considerably beyond $m_\pi/2$ and thus beyond the reach of both phase-quenched
and analytic-continuation methods.

We are performing simulations along lines of constant $\beta$ to determine the
equation-of-state for phase-quenched QCD to compare with that for full QCD
\cite{Ejiri:2005uv,Csikor:2004ik,Allton:2003vx}. At
low $\beta$s we are restricted to $\mu_I < m_\pi$. At high $\beta$s, where the
system is in the plasma phase for all $\mu_I$s, we can cover the whole range
of $\mu_I$. What can we learn from the resonance gas model \cite{Karsch:2003zq}
and chiral perturbation theory about this equation-of-state?

\section*{Acknowledgements}

DKS was supported in part by US Department of Energy contract 
DE-AC-02-06CH11357. JBK is supported in part by a National Science Foundation
grant NSF PHY03-04252. We thank Philippe de Forcrand and Frithjof Karsch for
helpful discussions. Simulations were performed on the Jazz cluster at Argonne,
Tungsten, Cobalt and Copper at NCSA, Bassi and Jacquard at NERSC, DataStar at
SDSC and PCs in Argonne's HEP Division.

\end{document}